\documentclass[5p,authoryear]{elsarticle}
\usepackage{hyperref}

\usepackage{multirow}

\usepackage{amsmath,amssymb}

\usepackage{fixltx2e}

\usepackage{color}
\newcommand{\numpy}{{\sf NumPy}}
\newcommand{\pyro}{{\sf pyro}}
\newcommand{\castro}{{\sf Castro}}
\newcommand{\flash}{{\sf Flash}}
\newcommand{\maestro}{{\sf Maestro}}

\journal{Astronomy and Computing}

\begin{document}

\begin{frontmatter}

\title{\pyro: A teaching code for computational astrophysical hydrodynamics}

\author{Michael Zingale\corref{cor}}
\address{Dept.\ of Physics and Astronomy, Stony Brook University, Stony Brook, NY, 11794-3800}
\cortext[cor]{Corresponding author. Tel.: 631-632-8225}
\ead{Michael.Zingale@stonybrook.edu}

\begin{abstract}
We describe \pyro: a simple, freely-available code to aid students in
learning the computational hydrodynamics methods widely used in
astrophysics.  \pyro\ is written with simplicity and learning in mind and
intended to allow students to experiment with various methods popular
in the field, including those for advection, compressible and incompressible
hydrodynamics, multigrid, and diffusion in a finite-volume framework.
We show some of the test problems from \pyro, describe its design
philosophy, and suggest extensions for students to build their
understanding of these methods.
\end{abstract}

\begin{keyword}
hydrodynamics \sep methods: numerical
\end{keyword}

\end{frontmatter}

\section{Introduction}

The majority of the algorithms used for astrophysical fluid flow are
first developed and described in journals devoted to applied math.
Traditionally, astrophysics students are not exposed to these journals
in their coursework, and their different target audience makes it
difficult for a new astrophysics graduate student to come up to speed
on the nuances of the methods.  It is also the case that the potential
audience of budding computational astrophysicist is sufficiently small
in a graduate year that regular course offerings have trouble meeting
the minimum class sizes imposed by a University.  Often the goals of
publicly-available production hydrodynamics codes,
e.g. \flash~\citep{flash} or \castro~\citep{almgren:2010}, are
not aligned with the needs of a teaching code.  In particular,
performance and breadth of options are favored over simplicity and
clarity.

Our experience from working on as a developer number of different
large simulation codes, including \flash~\citep{flash},
\castro, and \maestro~\citep{multilevel} and
with training students is that the
best way to learn computational methods for hydrodynamics is to code
them up yourself, or to make {\em substantial}\/ modifications to the
internals of existing code.  For the latter, it helps to have a simple
code as a starting point.  Here we describe \pyro\ (short for {\em
  py}\/thon hyd{\em ro}).  \pyro\ provides solvers for:
\begin{itemize}
    \addtolength{\itemsep}{-0.5\baselineskip}
\item linear advection
\item compressible hydrodynamics
\item elliptic PDEs via multigrid
\item implicit diffusion
\item incompressible hydrodynamics
\end{itemize}
All solvers are 2-d and second-order accurate in space and time.  We
chose 2-d because some of the key design issues in writing a solver
are not present in the simpler 1-d algorithms.  Furthermore, 2-d
offers sufficient complexity that the transition to 3-d is then
straightforward for a student.  Finally, 2-d allows for an exploration
of grid effects and instabilities in the solver that 1-d does not
allow.  \pyro\ is intended for self-study.  An accompanying set of
detailed lecture notes help explain the core methods and
experimentation is encouraged.  These notes have been used by
undergraduate researchers working with the author and are 
continually refined based on these interactions.

While there is a large variety of different methods for each
of these systems of equations, we pick a single method representative
of those used in astrophysics and implement it.  Given the choice
between clarity and performance, we take clarity.  Variations and
enhancements are left as exercises for students.

There are a number of excellent books that explain the basic theory of
numerical solution of PDEs, like \cite{leveque:2002} for finite-volume
methods and \cite{mgtutorial} for multigrid, but students also need
hands-on experience, to experiment, break, and tweak the algorithms.
Little details, like the number of ghost cells needed for different
parts of the algorithm are often not obvious to new students, so a
basic starting platform from which they can build on provides a good
introduction.  \pyro\ is written to help fill this need.  \pyro\ is
freely-available at \url{https://github.com/zingale/pyro2} with 
documentation provided at
\url{http://bender.astro.sunysb.edu/hydro_by_example/}

The core algorithms implemented in \pyro\ are not new---the new part
of \pyro\ is the focus on teaching the methods to the next generation
of students through a clean, robust implementation and hands-on
activities.  The purpose of this paper is to give an overview of the
algorithms \pyro\ provides, show some results from the various test
problems, demonstrating the validity of the methods we implement, and
provide ideas for extensions to help new students to the field build
their understanding.  This paper is complemented by detailed notes
describing the derivation and implementation of the various methods,
available on the \pyro\ website, and, of course, the freely-available
source code itself.  

\section{Design}

Python\footnote{\url{https://www.python.org/}} provides an attractive
platform for quickly testing out different ideas.  It is easy to use,
freely-available, and with the \numpy\footnote{\url{http://www.numpy.org/}} package, a powerful language
for manipulating arrays of data.  However, being an interpreted
language, the best performance is attained when you allow \numpy\ to
work on entire arrays of data instead of explicitly looping over the
individual elements.  There are some instances when the \numpy\ array
notation can look cumbersome, and hide from simple inspection the
differencing being done.  For example, consider constructing a
second-derivative as:
\begin{equation}
a^{\prime\prime}_i = \frac{a_{i+1} - 2 a_i + a_{i-1}}{\Delta x^2}
\end{equation}
Here the subscripts represent the index into a sequence of regularly
gridded data.
If we have a \numpy\ array {\tt a}, with appropriate ghost cells,
and use the integers {\tt lo} and {\tt hi} to refer to the first
and last valid cells, then we can write this in slice-notation
as:
\begin{verbatim}
d2a_dx2[lo:hi+1] = a[lo+1:hi+2] - \
               2.0*a[lo  :hi+1] + \
                   a[lo-1:hi]
d2a_dx2[:] = d2a_dx2[:]/dx**2
\end{verbatim}
This is efficient in python, but makes the underlying index notation
we are used to seeing in papers hidden (especially in 2-d).  For more complex
constructions, with nonlinear switches (e.g.\ {\tt if} conditions), 
the \numpy\ form can be complex.
To strike the right balance in terms of clarity and use of \numpy's
advanced features, we implement some kernels in Fortran, using {\sf
  f2py}\footnote{\url{http://cens.ioc.ee/projects/f2py2e/}; also part of \numpy} to interface with python.  Wherever Fortran is used, we
enforce the following design rule: the Fortran functions must be
completely self-contained, with all information coming through the
interface. No external dependencies are allowed. Each \pyro\ module
will have (at most) a single Fortran file and can be compiled into a
library via a single {\sf f2py} command line invocation.

There are two fundamental classes in \pyro\ that manage the data.  The
{\tt Grid2d} class describes the grid, providing the basic coordinate
information and the {\tt CellCenterData2d} class describes the data
that lives on the grid.  Building a {\tt CellCenterData2d} object
takes a {\tt Grid2d} object at initialization and has methods to
register variables that live on the grid.  Each variable can have its
own boundary condition types and methods exist for getting access to
a single variable, filling ghost cells, restricting and prolonging a
data to a new grid (used by the multigrid solver), printing data to
the screen (useful only for small grids), and writing the data to
disk.

Each solver is given its own directory, with {\tt problems/} and {\tt
  tests/} sub-directories.  The former holds the initial condition
routines and default parameters for each of the problems known to that
solver.  The latter stores benchmark output and is used for the
built-in regression testing.

\begin{table*}[t]
\centering
\renewcommand{\arraystretch}{1.2}    
\begin{tabular}{llp{4.0in}}
\hline
{\bf solver} & {\bf problem} & {\bf problem description} \\
\hline
\multirow{2}{*}{\tt advection} & {\tt smooth} & advect a smooth Gaussian profile \\
                               & {\tt tophat} & advect a discontinuous tophat profile \\
\hline
\multirow{6}{*}{\tt compressible} & {\tt bubble} & a buoyant bubble in a stratified atmosphere \\
                                  & {\tt kh} & setup a shear layer to drive Kelvin-Helmholtz instabilities \\
                                  & {\tt quad} & 2-d Riemann problem based on \citet{schulz-rinne:1993}  \\
                                  & {\tt rt}  & a simple Rayleigh-Taylor instability\\
                                  & {\tt sedov} & the classic Sedov-Taylor blast wave \\
                                  & {\tt sod} & the Sod shock tube problem  \\
\hline
{\tt diffusion}& {\tt gaussian} & diffuse an initial Gaussian profile \\
\hline
\multirow{2}{*}{\tt incompressible} & {\tt converge} & A simple incompressible problem with known analytic solution.  \\
                     & {\tt shear} & a doubly-periodic shear layer \\
\hline
\end{tabular}
\caption{\label{tab:problems} Solvers and their distributed problems}
\renewcommand{\arraystretch}{1.0}    
\end{table*}

One of the concepts that comes out of the code is how similar the
solution methodology is for the different PDE systems.  The grid
requirements/data locations are the same, the boundary conditions types are
analogous, and even the overall flowchart of the main driver is the
same.  For all time-dependent solvers (i.e., excepting multigrid)
the basic flowchart is:
\begin{itemize}
  \addtolength{\itemsep}{-0.5\baselineskip}
\item parse runtime parameters
\item setup the grid 
\item set the initial conditions for the data on the grid
\item do any necessary pre-evolution initialization
\item evolve while $t < \mathtt{tmax}$ or $n < \mathtt{max\_steps}$:
\begin{itemize}
  \addtolength{\itemsep}{-0.25\baselineskip}
  \item fill boundary conditions 
  \item get the timestep 
  \item evolve for a single timestep 
  \item t = t + dt;\, n = n + 1
  \item output 
  \item visualization 
\end{itemize}
\end{itemize}

This allows us to have a single driver for all the solvers.  \pyro\ is run
as:
\begin{quote}
{\tt ./pyro.py} [{\em options}] {\em solver} {\em problem} {\em infile} [{\em runtime-options}]
\end{quote}
where {\em solver} is one of {\tt advection}, {\tt compressible}, {\tt
  diffusion}, or {\tt incompressible}.  The {\em problem} gives the
name of the problem whose initial conditions we use---these vary by
solver (see Table~\ref{tab:problems}).  Finally the {\em infile}
overrides any default runtime parameters for the run.  Runtime
parameters are defined both in the main \pyro\ directory and for each
solver and problem in plain text files that are parsed at runtime.
Optionally, runtime parameters defaults can also be overridden at the
end of the commandline.  The collection of runtime parameters is managed
by the {\tt RuntimeParameters} class.

The interaction with the different solvers is done through each
solver's {\tt Simulation} class which holds the simulation's {\tt RuntimeParameters} object, the {\tt CellCenterData2d}
data, and some timer
information (for profiling).  The class provides the following methods:
\begin{itemize}
  \addtolength{\itemsep}{-0.5\baselineskip}
\item {\tt initialize()}: set up the grid and solution variables.

\item {\tt timestep()}: return the timestep for evolving the system.

\item {\tt preevolve()}: do any initialization to the fluid state that is 
necessary before the main evolution. Not every solver will need something here.

\item {\tt evolve()}: advance the system of equations through a single 
timestep.

\item {\tt dovis()}: perform visualization of the current solution.

\item {\tt finalize()}: any final clean-ups, printing of analysis hints.
\end{itemize}

I/O is done simply using the python {\tt pickle} module on the main
data object\footnote{ {\tt pickle} is a part of the standard python
  library that serializes an object into a sequence of bytes that can
  be written to disk.  See
  \url{https://docs.python.org/2/library/pickle.html} }.  Finally, by
default, runtime visualization is enabled using the {\sf
  matplotlib}\footnote{\url{http://matplotlib.org/}} plotting library.
The output is updated each timestep to allow students to see the
progression of their simulations as they run.  This is very useful for
seeing the effects of boundary conditions and different choices of
initial conditions as well as for chasing down bugs.

\subsection{Software engineering ideas}

Astronomy students benefit from learning basic ``safe practices'' from
software engineering~\citep{ferland:2002,wilson:2012,turk:2013}.
\pyro\ facilitates this in three ways.  First, pyro is open source,
released under a BSD-3 license.  Sharing of code helps find bugs, as
more eyes are now on the source.  Furthermore, open code is
increasingly being seen as part of the scientific peer-review
proces~\citep{shamir:2013}.

Secondly, support for regression
testing is built into pyro.  In its simplest form, regression testing means
comparing the current solution to a known benchmark solution, and
looking for differences.  When \pyro\ is run with the {\tt
  --make\_benchmark} option, it will automatically store a benchmark
solution for the problem in the solver's {\tt tests/} sub-directory.
When run with the {\tt --compare\_benchmark} option, the current
results will be compared zone-by-zone with the stored benchmark, and
any differences will be reported.  We note that comparing bit-for-bit
in each zone means that we may not find agreement when comparing
across different platforms because of floating point differences.  An
alternative is check for agreement in each zone to within some
tolerance.  Both approaches have value.  We prefer the exact
comparison and it is what is used in the regression suites for the
production codes we have worked with---e.g., this is done by the {\tt sfocu}
comparison tool in the {\tt flashTest} regression
framework \citep{flashdoc}.  If there is some round-off
level difference in output, checking to within a tolerance may not
catch it, but the exact comparison will.  The developer can then check
to see if a code change is responsible for floating point differences
or whether some underlying change to the software enviroment is at
play.  In either scenario, the developer is made aware of a difference
in the output that comparison to within some tolerance would miss.
The downside is that unique benchmarks may need to be made on each
platform.

Finally, as is increasingly the case in astrophysics (see e.g.\ the
\castro\ code or {\tt yt} \citealt{yt}) it is obtained via a
distributed version control system (git in our case, recently migrated
to {\sf GitHub}\footnote{\url{https://github.com}}), immediately
immersing students in version control---each student's clone of the
\pyro\ repo acts as its own git repo that they can interact with
directly.  We have taught several graduate classes where a basic
introduction to programming practices was given, and we found that
many students were simply unaware of the idea of version control.
Once the concept was introduced, we have seen many students start to
use it for their own projects.

\section{Algorithm Summaries and Tests}

Here we give an overview of the methods \pyro\ implements and show some
verification tests.

\begin{figure*}[t]
\centering
\includegraphics[width=0.9\linewidth]{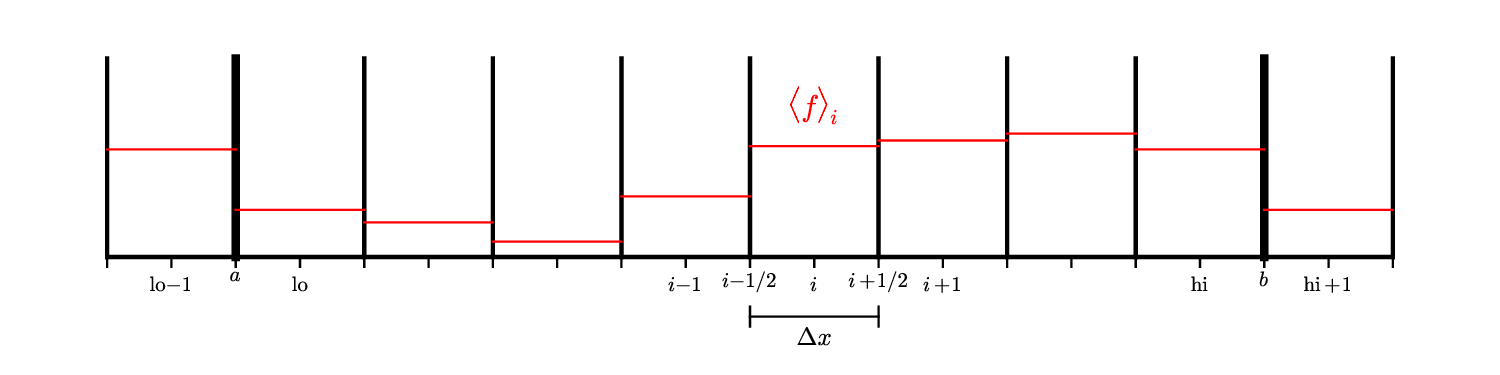} 
\caption{\label{fig:grids} A (1-d) finite-volume grid.  The domain
  boundaries are indicated by the thick lines and a single ghost cell
  is shown.  $\mathrm{lo}$ and $\mathrm{hi}$ indicate the first and
  last valid cells.  $a$ and $b$ mark the locations of the left and
  right boundaries.}
\end{figure*}

When solving a system of PDEs, continuous derivatives are replaced by
their discretized counterparts, either by making reference to an
underlying grid, or by using a collection of particles to represent
the functional distribution.  Within astrophysics, both grid base and
particle-base methods are popular, and each has their own strengths
and weaknesses (see e.g.~\citealt{agertz:2007,hubber:2013} and
references therein). We focus here on structured grids---logically
Cartesian grids where any zone can be references by a single integer
in each dimension.  Structured grids are popular because the
geometries of our domains are generally not complex, and because
structured grids can make good use of modern cache-based
architectures.  Both finite-difference and finite-volume methods are
structured-grid discretizations popular in astrophysics.  A
finite-volume method represents the data as an average within a volume
(or zone) while a finite-difference grid stores the function value at
a specific point.  These differences arise from whether we choose to
work with the integral form (finite-volume) or differential form
(finite-difference) of our system of equations.  We note that a
cell-centered finite-difference grid is equivalent to a finite-volume
grid to second-order in $\Delta x$.  We choose the finite-volume grid
here, and use the convention that zone centers are indicated by an
integer, $i$, while the interfaces are marked by a half-integer
($i-1/2$ on the left and $i+1/2$ on the right).
Figure~\ref{fig:grids} illustrates the grid in 1-d.

\subsection{Advection}

The linear advection equation represents the simplest hyperbolic PDE:
\begin{equation}
a_t + u a_x + v a_y = 0
\end{equation}
The solution is trivial---any initial function profile simply advects
unchanged with a velocity $u \hat{x} + v \hat{y}$.  This makes
advection an excellent test bed of numerical methods.

Advection also provides a good path toward understanding how to extend
to multi-dimensions.  There are two approaches here: dimensional
splitting and unsplit reconstruction.  In dimensional splitting, the
fluxes at the interfaces are constructed without any knowledge of the
flow in the transverse direction.  Each directional update operates on
the state left behind by the previous update and the order of
directions is alternated to give second-order
accuracy~\citep{strang:1968}.  This is the easiest way to extend from
one-dimension to multi-dimensions, because you can use the 1-d
methodology largely unchanged.  However, split methods do not preserve
symmetries as well (see, e.g.\ \citealt{almgren:2010}).  In an unsplit
reconstruction, each interface state explicitly sees the change
carried in the transverse direction, and the state is updated by the
fluxes in each direction all at once.

\begin{figure*}[t]
\centering
\includegraphics[width=0.45\linewidth]{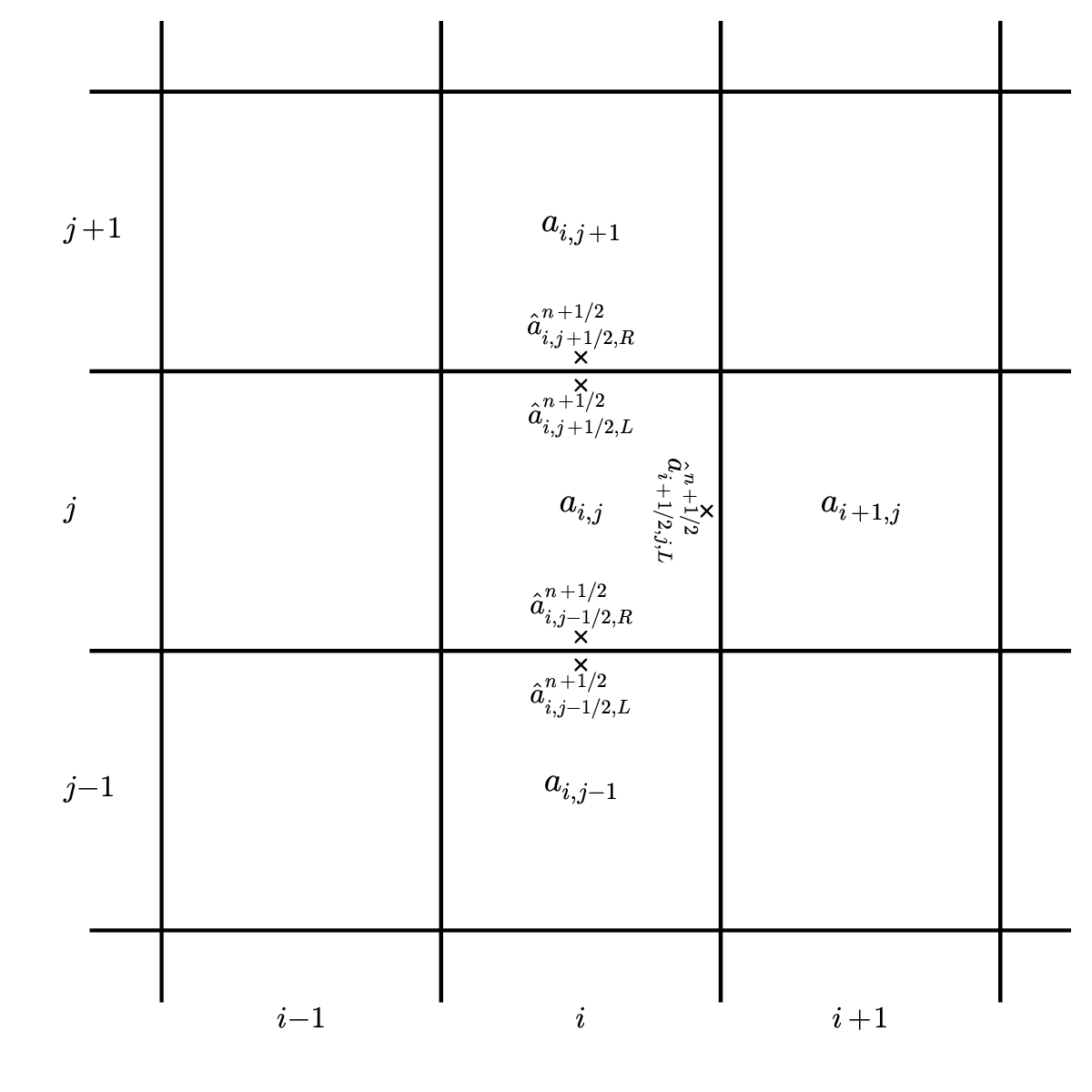}
\includegraphics[width=0.45\linewidth]{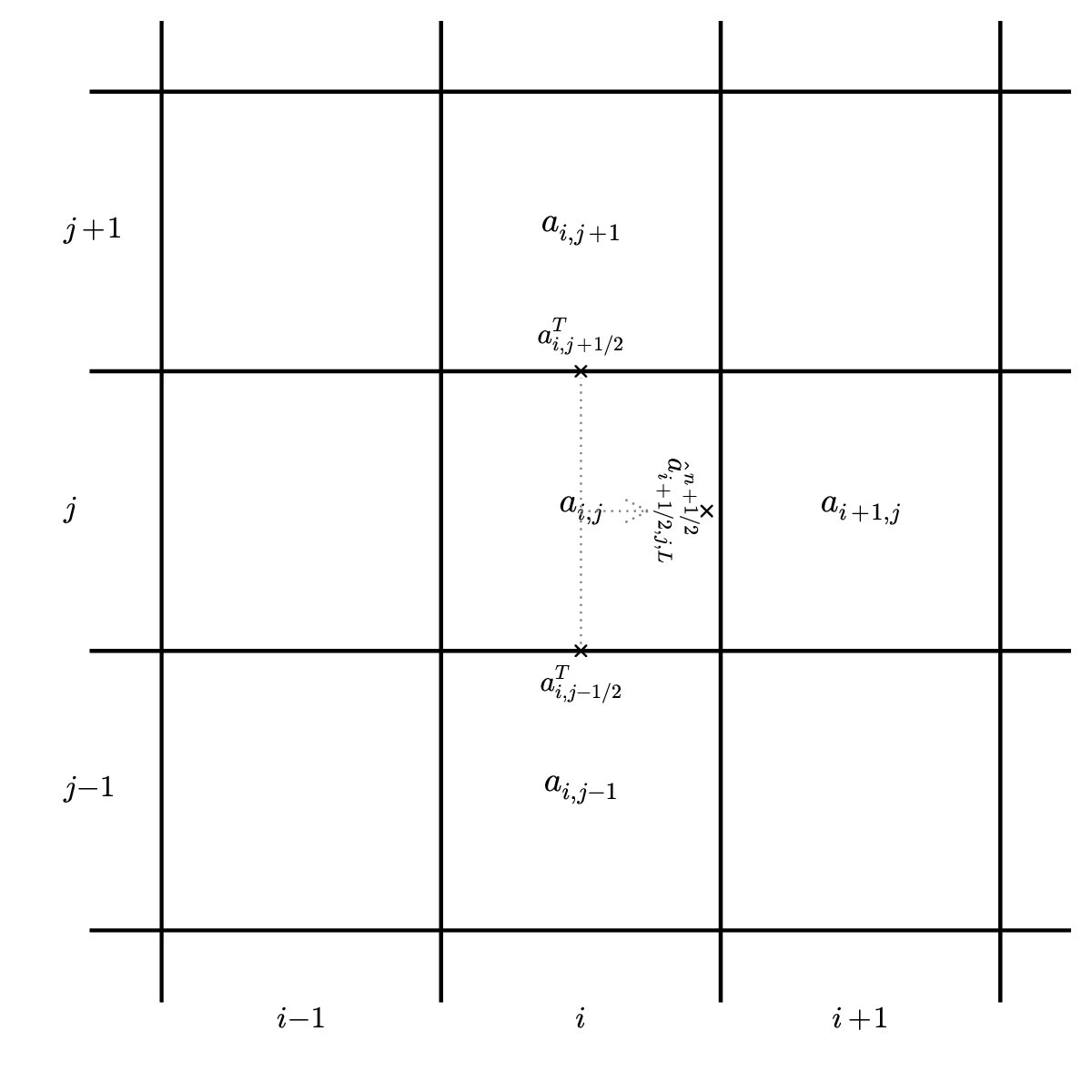}
\caption{\label{fig:unsplitstates} The construction of the
  $a_{i+1/2,j,L}^{n+1/2}$ state.  Left: first we compute the
  $\hat{a}$'s---here we show all of the $\hat{a}$'s that will be used
  in computing the full left interface state at $(i+1/2,j)$.  Right:
  after the transverse Riemann solves, we have the two transverse
  states ($a^T_{i,j+1/2}$ and $a^T_{i,j-1/2}$) that will be
  differenced and used to correct $\hat{a}_{i+1/2,j,L}^{n+1/2}$
  (illustrated by the dotted lines) to make $a_{i+1/2,j,L}^{n+1/2}$.}
\end{figure*}

We follow the unsplit CTU method~\citep{colella:1990} for advection.
We summarize this method in a little detail below because the same
procedure comes into play again for our compressible and
incompressible solvers.  The basic idea is simple---Taylor expand in
space and time to get the time-centered interface state and use this
to evaluate the fluxes through the interfaces.  For example, the left
state at the $i+1/2,j$ interface is built starting with the data in
cell $i,j$ as
\begin{equation}
a_{i+1/2,j,L}^{n+1/2} = a_{i,j}^n +
   \frac{1}{2} \left ( 1 - \frac{\Delta t}{\Delta x} u \right )
   \left . \overline{\Delta a} \right |_{i,j} - \frac{\Delta t}{2} v \left .\frac{\partial a}{\partial y} \right |_{i,j} \label{eq:advstate}
\end{equation}
The last term here is the transverse flux difference and captures the
change in $a$ in the transverse direction.  Without this term, if we
advect diagonally, we would not `see' the upwind state.  The
$\overline{\Delta a}$ term is the limited slope of $a$---limiting
ensures that we don't have any over- or under-shoots when we advect
(although see \cite{BDS} for details about limiting in
multi-dimensions).  A number of different choices of limiters are
described in the literature.  We use the 4th-order MC limiter from
\citet{colella:1985}.  
We rewrite this as:
\begin{equation}
a_{i+1/2,j,L}^{n+1/2} = \hat{a}_{i+1/2,j,L}^{n+1/2}
   - \frac{\Delta t}{2} v \left .\frac{\partial a}{\partial y} \right |_{i,j}
\end{equation}
where the $\hat{a}_{i+1/2,j,L}^{n+1/2}$ state represents the
prediction to the interface without regard to the transverse term.
The basic idea of the CTU method is to first construct these `hat'
states on all interfaces, then solve the Riemann problem at each
interface to find the `transverse' state, $a^T_{i+1/2,j}$---this is
the unique state on each interface.  So far however, we did not
consider the transverse term in Eq.~\ref{eq:advstate}.  This
transverse term is constructed using the $a^T$ edge states and added
to the $\hat{a}$ states giving us the full interface state
$a_{i+1/2,j,L}^{n+1/2}$.  This is illustrated graphically in
Figure~\ref{fig:unsplitstates}.

A similar reconstruction gives the predicted state just to the right
of the interface, $a_{i+1/2,j,R}^{n+1/2}$, building from the data in
the $i+1,j$ cell.  The final state on the interface,
$a_{i+1/2,j}^{n+1/2}$ is attained by calling the Riemann solver again.
For advection, the Riemann solve is simply upwinding, but for other
systems it is more complex.  The states in the $y$-direction are
computed analogously.  This then allows us to increment our solution:
\begin{align}
 \frac{a_{i,j}^{n+1} - a_{i,j}^n}{\Delta t} =
  &- \frac{ (u a)_{i+1/2,j}^{n+1/2} - (u a)_{i-1/2,j}^{n+1/2} }{\Delta x} \nonumber \\
  &- \frac{ (v a)_{i,j+1/2}^{n+1/2} - (v a)_{i,j-1/2}^{n+1/2} }{\Delta y}
\label{eq:update2du}
\end{align}

The timestep constraint is simply 
\begin{equation}
\Delta t = C \min \left \{ \frac{\Delta x}{u}, \frac{\Delta y}{v} \right \}
\end{equation}
where $C \le 1$ is the CFL number.

\subsubsection{Smooth advection test} 

\begin{figure*}[t]
\centering
\includegraphics[width=0.5\linewidth]{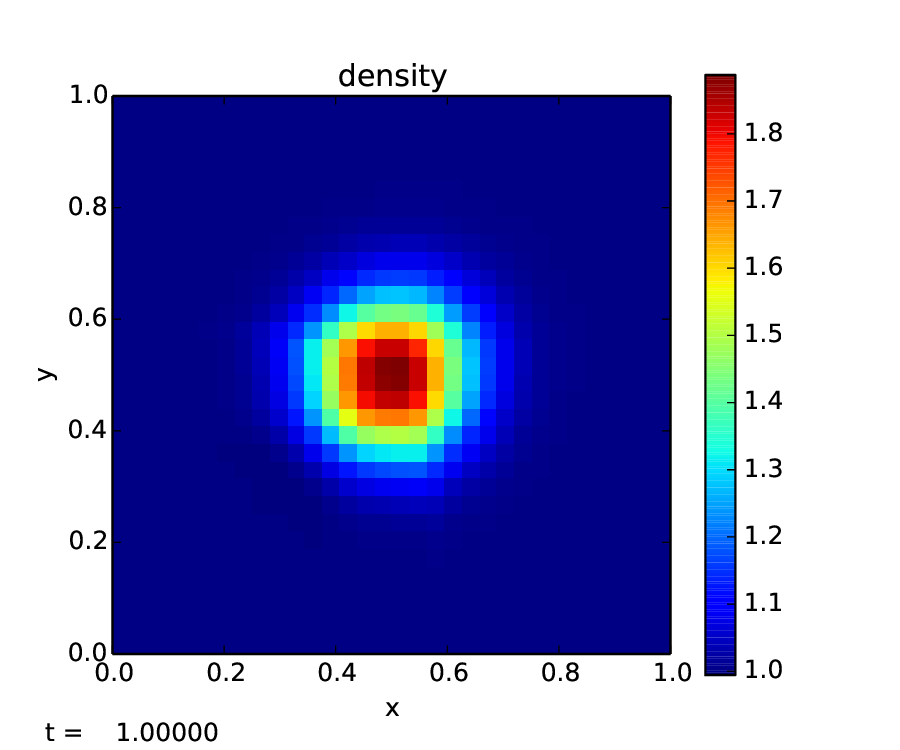}
\includegraphics[width=0.4\linewidth]{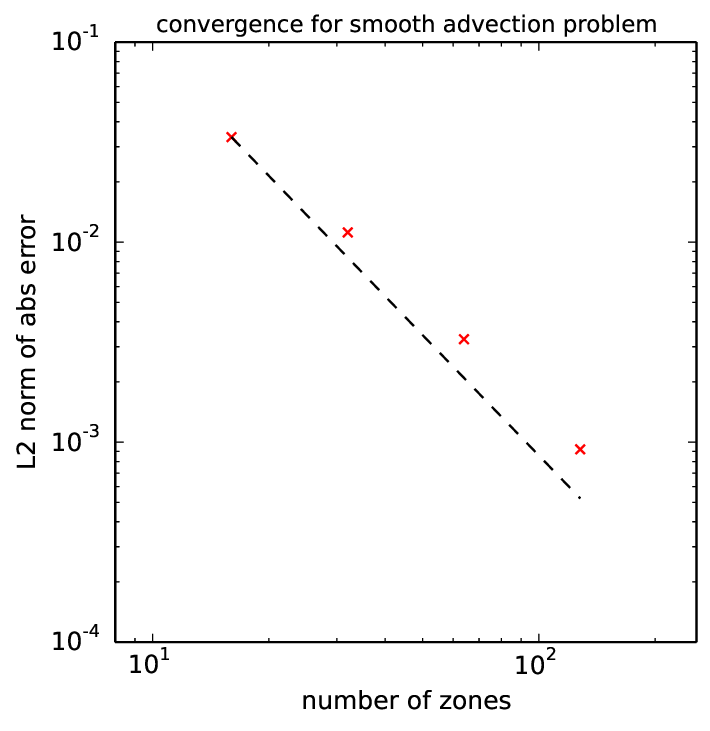}
\caption{\label{fig:advect} Linear advection of a Gaussian profile.  Left shows the
  profile on a $32^2$ grid after advecting diagonally for a period.
  Right shows the convergence of the solver, with the dotted line
  representing $\mathcal{O}(\Delta x^2)$ scaling.  This example
  was run with: {\tt ./pyro.py advection smooth inputs.smooth}\/.
  The different resolutions were run by changing {\tt mesh.nx} and
  {\tt mesh.ny}.
  The errors for the plot on the right were computed with {\tt analysis/smooth\_error.py}.}
\end{figure*}

The simplest test is to advect a smooth distribution diagonally across
our grid for one period.  Since the profile should be unchanged, we
can define the error simply as the difference between the final and
initial profile, and look at the norm of the error to assess the
convergence.  A smooth problem is chosen so as to reduce the effects
of the limiters---we choose a Gaussian:
\begin{equation}
a(x,y) = 1 + e^{-A [ (x - x_c)^2 + (y - y_c)^2 ]}
\end{equation}
where $(x_c,y_c)$ are the coordinates of the center of the domain and
$A$ is a width parameter.  We choose $A = 60$.
Figure~\ref{fig:advect} shows the solution after 1 period with $u = v
= 1$, a CFL number of 0.8, and for one resolution ($32^2$) along with
a convergence plot of the error as a function of resolution, showing
that we are nearly second-order accurate.  The departure from perfect
second-order convergence arises from the use of limiters.

\subsubsection{Explorations for students}
With this basic solver, there are a number of exercises/extensions
that students can perform to improve their understanding:
\begin{itemize}
    \addtolength{\itemsep}{-0.5\baselineskip}
\item Compare convergence with limiting to no limiting.
\item How does the solution change if the transverse flux difference
  is left out?  
\item Implement a dimensionally split version and compare to
  the unsplit. 
\item Convert the solver to the inviscid Burger's equation ($u_t + u
  u_x = 0$) and look at shocks and rarefactions.
\end{itemize}

\subsection{Compressible hydrodynamics}

Compressible hydrodynamics is described by the Euler equations:
\begin{eqnarray}
\rho_t + \nabla \cdot (\rho U) &=& 0 \\
(\rho U)_t + \nabla \cdot (\rho U U) + \nabla p &=& \rho g \\
(\rho E)_t + \nabla \cdot (\rho U E + U p) &=& \rho U \cdot g
\end{eqnarray}
Here, $\rho$ is the density, $U$ is the velocity vector, $p$ is the
pressure, and $E$ is the total specific energy, related to the specific
internal energy, $e$, via
\begin{equation}
E = e + \frac{1}{2} U^2
\end{equation}
We include a gravitational source, with $g$ the constant gravitational
acceleration.
We need an equation of state to close the system.  We assume a gamma-law:
\begin{equation}
p = \rho e (\gamma - 1)
\end{equation}

The solution procedure we adopt is the unsplit piecewise-linear method
described in \citet{colella:1990}.  This same algorithm is also one of
the hydrodynamics options in the \castro\ code~\citep{almgren:2010}.
While formally less accurate than the widely-used piecewise parabolic
method, PPM~\citep{colellawoodward:1984}, piecewise linear
reconstruction is more approachable for new students, and upgrading to
PPM is straightforward once the details are understood (see
\citealt{millercolella:2002} for a good discussion).

The algorithm follows the advection update closely.  Again, interface
states are constructed by doing a Taylor expansion to the half-time,
interface state.  Now however, the normal part of the prediction is
done in terms of the primitive variables, $q = (\rho, u, v, p)^\intercal$,
instead of the conserved variables.  A characteristic projection is
done on the state and only the jumps moving toward the interface are
added to the interface state.  These preliminary interface states are
converted back to the conserved variables and the transverse flux
difference is added.  An initial construction of the normal interface
states is used to construct the fluxes for the transverse difference.
Our implementation follows \citet{colella:1990}.  We also include
the flattening of the profiles near shocks and artificial viscosity
from that paper.

One confusing aspect of these methods for new students is the 
characteristic projection.  Writing the primitive variable system
as $q_t + A q_x = 0$, the jump in a primitive variable,
$\Delta q$ can be expressed in terms of the left and right
eigenvectors, $l$ and $r$ of $A$ as:
\begin{equation}
\Delta q = \sum_\nu (l^{(\nu)} \cdot \Delta q) r^{(\nu)}
\end{equation}
\citet{leveque:2002} provides an excellent introduction to this concept.
Typically in this sum, we only include the terms where the waves are
moving toward the interface of interest (determined by the sign of the
eigenvalues).  Most papers write only the analytic result of
multiplying out $(l^{(\nu)} \cdot \Delta q)$ (e.g.\ see the $\beta$'s
in Eqs. 3.6, 3.7 in \citealt{colellawoodward:1984}).  For clarity, and to enable
exploration, in \pyro, we explicitly construct the left and right
eigenvectors and multiply them through in the code, to illustrate
exactly what the solution procedure is doing.

The Riemann problem for compressible hydrodynamics is considerably
more complex than linear advection.  There are 3 waves that result
from the system (corresponding to the $u-c$, $u$, and $u+c$
eigenvalues of $A$), and each wave, $\nu$, carries a jump in the state
proportional to $r^{(\nu)}$.  The solution to the Riemann problem
looks at wave structure to determine the solution in-between the
various waves and evaluates the wavespeeds to determine which state is
on the interface.  This state is then used to construct the fluxes
through the interface.  Because of the expense of the full Riemann
solve, approximate Riemann solvers are often used.  We use the Riemann
solver described in \cite{almgren:2010}, and alternately the HLLC
method from \citet{toro:1997}.


\subsubsection{Sod shock tube}

\begin{figure}[t]
\centering
\includegraphics[width=0.9\linewidth]{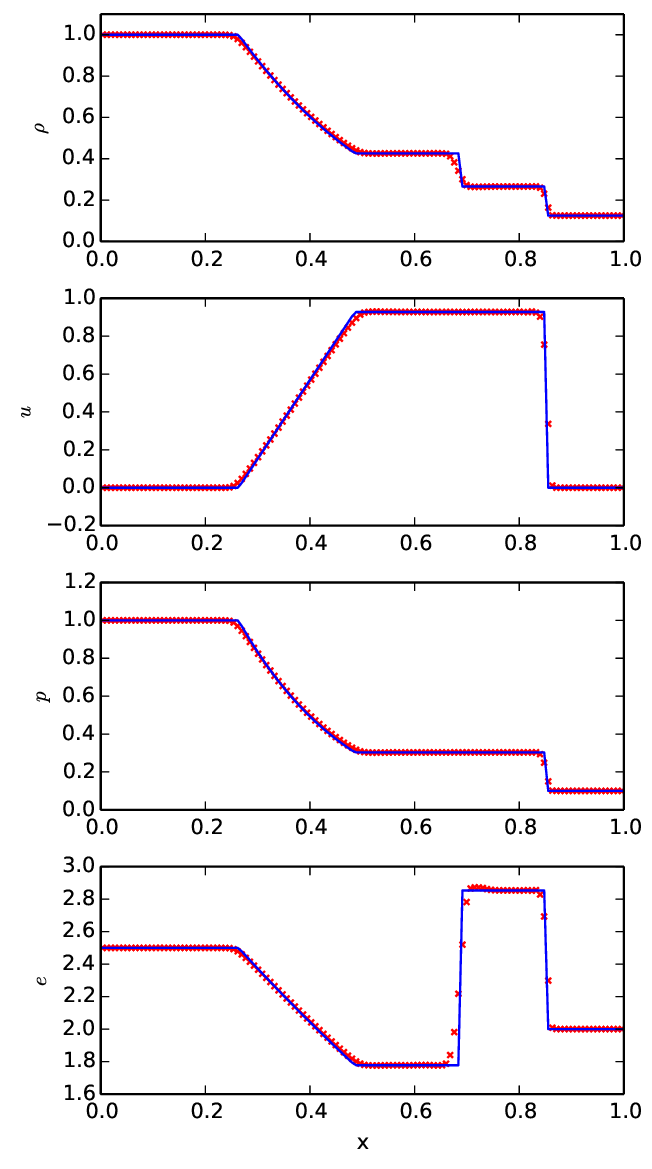}
\caption{\label{fig:sod} Sod problem with 128 zones compared to the
  exact solution at t = 0.2.  We show a slice through the center of
  the domain along $x$.  This example was run with: {\tt ./pyro.py
    compressible sod inputs.sod.x }, and the comparison to the analytic
    data was done with {\tt analysis/sod\_compare.py}.}
\end{figure}

The Sod shock tube~\citep{sod:1978} illustrates all three types of
hydrodynamic waves: a rarefaction, contact, and shock wave.  This
problem is a standard test of hydrodynamics solvers because exact
solutions are possible.  We compare to the result from the exact
Riemann solver in \citet{toro:1997}.

Figure~\ref{fig:sod} shows the results from a simulation on a
$128\times 10$ grid with the initial discontinuity in the
$x$-direction.  The CFL number was 0.8 and $\gamma = 1.4$.  The shock
is very steep, the contact is smeared out a bit (there is no
self-steepening mechanism for contacts, so this is commonly seen).
Overall the solution matches the analytic profile well.  The 
discontinuities in the solution mean that taking the norm of
the error with the exact solution has little value.

\subsubsection{Sedov-Taylor blast wave}

\begin{figure*}[t]
\centering
\includegraphics[width=0.9\linewidth]{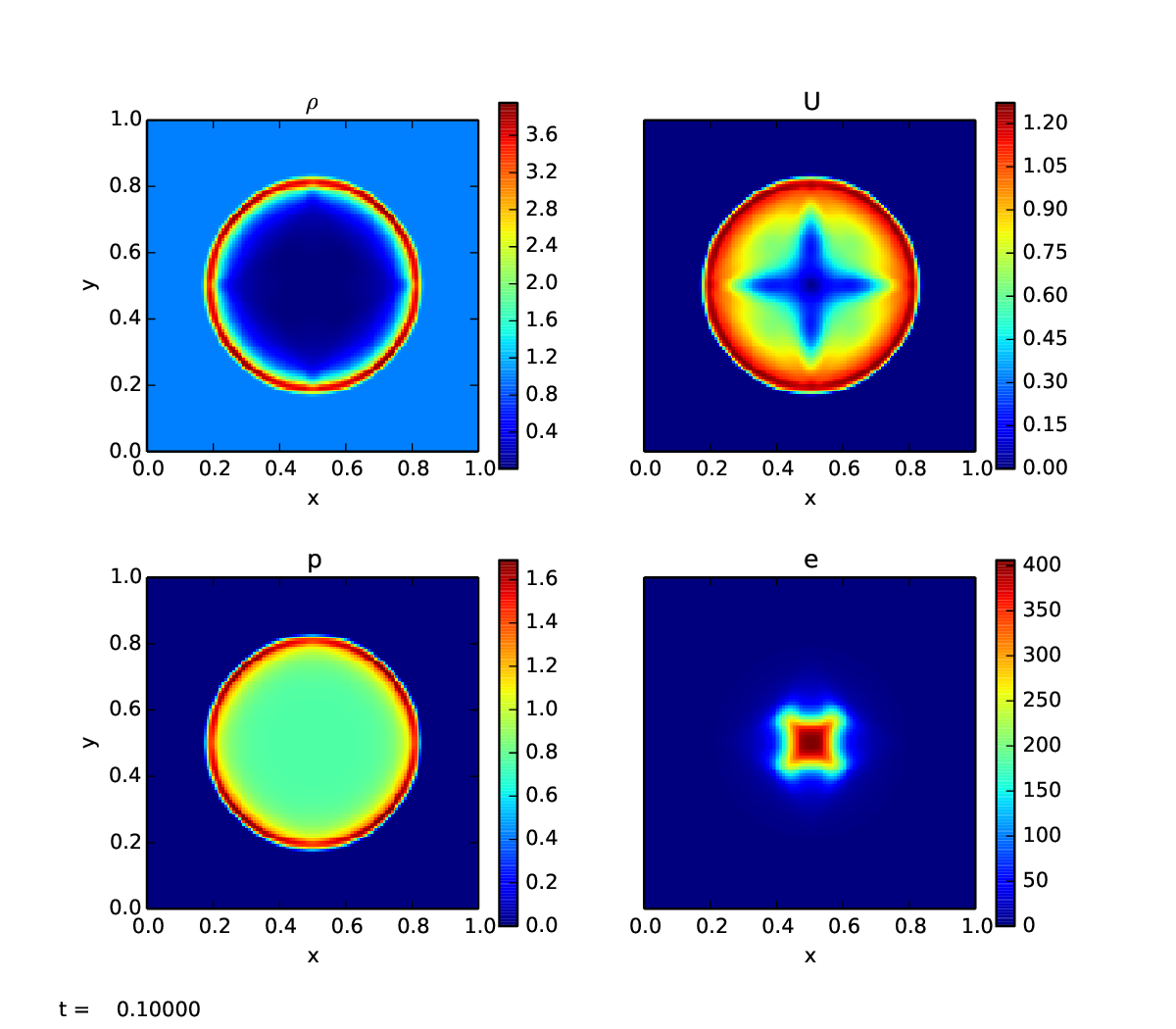}
\caption{\label{fig:sedov2d} Solution to the Sedov problem at $t = 0.1$.
This example was run with: {\tt ./pyro.py compressible sedov inputs.sedov}}
\end{figure*}

\begin{figure}[t]
\centering
\includegraphics[width=0.9\linewidth]{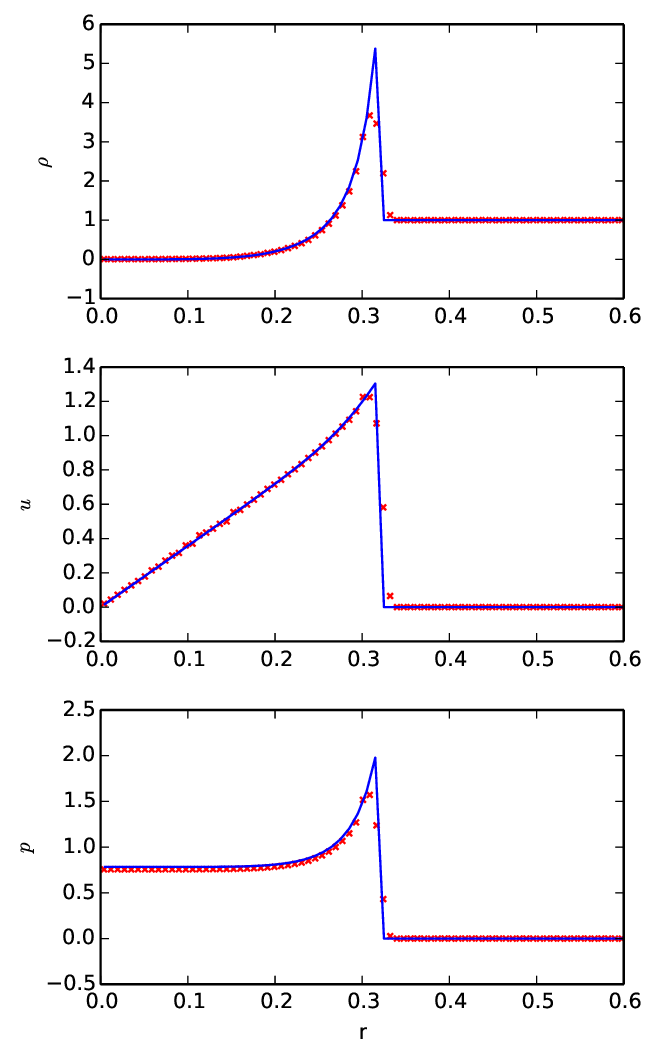}
\caption{\label{fig:sedov} Angle-averaged profiles for the Sedov
  problem for a $128^2$ simulation at $t = 0.1$. The solid line is the
  exact solution.  This figure was created with the {\tt analysis/sedov\_compare.py} script.}
\end{figure}

The Sedov-Taylor blast wave is a good test of the
sym\-metry-preservation of the method.  A large amount of energy,
$\mathcal{E}$, is placed at a point in a uniform medium.  The blast
wave should stay circular (in 2-d), with the density evacuating and
the pressure reaching a constant at the center.  An analytic solution
was worked out for this problem \citep{sedov:1959}.  A good
description of the solution is given in \citet{timmes_sedov_code}---we
compare to the solution from that paper.

A difficulty with the Sedov problem on a 2-d Cartesian grid is that
the point that the energy is deposited to will be square on the grid,
leading to some grid effects.  We take the standard approach of
converting the energy into a pressure contained in a circular region:
\begin{equation}
p = (\gamma - 1) \frac{\mathcal{E}}{\pi r_\mathrm{init}^2}
\end{equation}
To further reduce grid effects, we sub-divide each zone into $4^2$
sub-zones and test whether each sub-zone falls inside the perturbed
radius, and average over the sub-zones to get a single pressure for
the zone.  We choose $\mathcal{E} = 1.0$, $r_\mathrm{init} = 0.01$,
and $\gamma = 1.4$.  We run on a $128^2$ grid with a CFL number of
0.8.  Figure~\ref{fig:sedov2d} shows the various fluid quantities.  We
see a nice circular blast wave, but some grid effects are seen aligned
with the coordinate axes.  The angle-averaged data (profiles as a
function of radius) are shown in Figure~\ref{fig:sedov}.  We see very
good agreement with the exact solution.

\subsubsection{Rayleigh-Taylor instability}

The Rayleigh-Taylor instability places a dense fluid over a lighter fluid
in a gravitational field.  Given an initial perturbation, the dense fluid
drops down and the lighter fluid buoyantly rises upward.  Our initial
conditions are:
\begin{equation}
    \rho = \left \{ \begin{array}{cc} 
              \rho_1 
                & \mathrm{if~} y < y_c \\
              \rho_2 
                & \mathrm{if~} y \ge y_c  
     \end{array}  \right .
\end{equation}
The pressure is given by hydrostatic equilibrium, which for constant
$g$ integrates easily giving:
\begin{equation}
    p = \left \{ \begin{array}{cc} 
       p_0 + \rho_1 g y  
         & \mathrm{if~} y < y_c \\
       p_0 + \rho_1 g y_c + \rho_2 g(y - y_c) 
         & \mathrm{if~} y \ge y_c
     \end{array} \right .
\end{equation}
The perturbation is given in the $y$-velocity:
\begin{equation}
  v = A \cos(2\pi x/L_x) * e^{-(y-y_c)^2/\sigma^2}
\end{equation}

\begin{figure*}[t]
\centering
\includegraphics[width=0.9\linewidth]{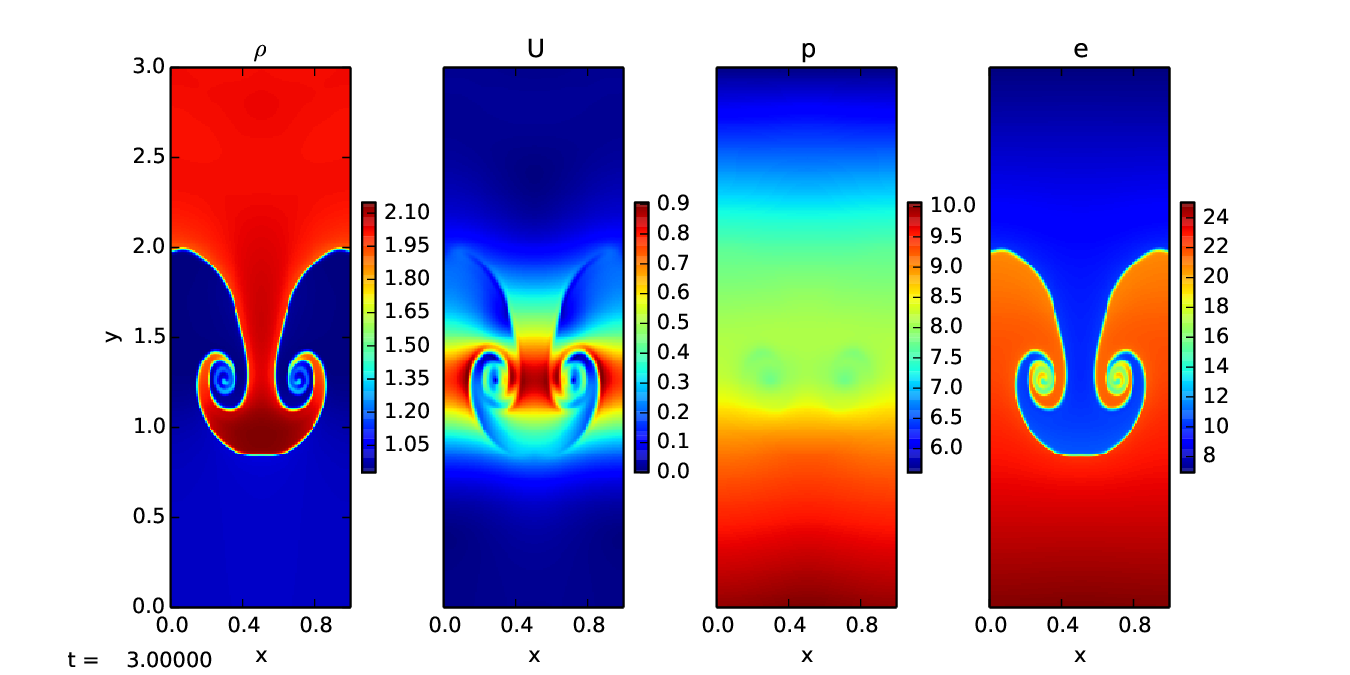}
\caption{\label{fig:rt} A single-mode Rayleigh-Taylor instability
  calculation with a 2:1 density ratio at $t = 3.0$.  This was run
  with {\tt ./pyro.py compressible rt inputs.rt}}
\end{figure*}

Periodic boundary conditions are used in the horizontal direction.  We
use hydrostatic boundary conditions in the vertical direction.  This
is one example of where the treatment of the boundary is
solver-dependent, so the compressible solver provides a method that is
called from the main mesh module to extend the functionality of the
boundary filling.  For this boundary condition, the density and $x$-
and $y$-momenta are simply copied from the last valid zone inside the domain
(a zero-gradient).  The energy is filled in the ghost cells by first
constructing the pressure in the zone just inside the boundary and then using
hydrostatic equilibrium to integrate this pressure into the ghost cells.  Since
the density and gravity are constant, this integration is trivial.  This 
pressure is then used to construct the total energy in the ghost cells.  This
method is similar to that from \citet{hse}.

Figure~\ref{fig:rt} shows the results for a $128\times 384$ zone grid
with a CFL of 0.8.  We see that this single mode RT is very symmetric,
and in good agreement with the results shown in \cite{almgren:2010}.

\subsubsection{Explorations for students}

After understanding the basic algorithm, there are a number of
extensions that can be done to build a deeper understanding of the
method.

\begin{itemize}
    \addtolength{\itemsep}{-0.5\baselineskip}\item Compare the solutions with and without limiting.

\item Limit on the characteristic variables instead of the primitive
  variables (see \citealt{athena}). 

\item Add passively advected species to the solver.

\item Add an external heating term to the equations.

\item Add 2-d axisymmetric coordinates (r-z) to the solver.

\item Swap the piecewise linear reconstruction for piecewise parabolic
  (PPM). 

\item Add different Riemann solvers to the algorithm.

\end{itemize}

\subsection{Multigrid}

\begin{figure}
\centering
\includegraphics[width=0.9\linewidth]{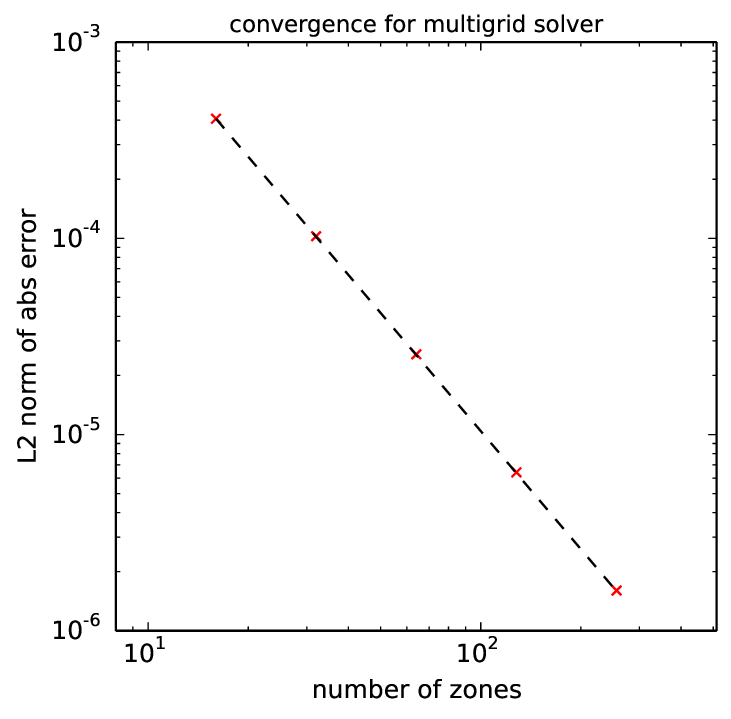}
\caption{\label{fig:mg} L2 norm of the error for the multigrid test
  problem, showing second-order accuracy.  This example was run
  at several resolutions using the {\tt mg\_test.py} script in the {\tt multigrid} module---that script reports the error at the end of execution. }
\end{figure}

Multigrid is a popular technique for solving elliptic problems in
simulation codes (a widely-used example is the Poisson problem for the
gravitational potential).  The \pyro\ multigrid solver solves a
constant-coefficient Helmholtz equation of the form:
\begin{equation}
\label{eq:helmholtz}
(\alpha - \beta \nabla^2) \phi = f
\end{equation}
The choice $(\alpha, \beta) = (0, -1)$ gives the classic Poisson equation.

The text of \cite{mgtutorial} provides an excellent introduction to
multigrid techniques.  Most of it, however, is geared toward
finite-difference grids, where solution values exist explicitly on the
boundary.  For cell-centered/finite-volume grids (see
Figure~\ref{fig:grids}), the implementation of the restriction and
prolongation operations differs, as does the enforcement of the
boundary conditions.  In fact, boundary conditions for cell-centered
grids is probably the most common place that mistakes are made.  A
good way to test this all is to get pure relaxation working first, and
converging to second-order.  Multigrid simply accelerates the
convergence of the relaxation, so if the relaxation routine is not
right, then there is no point going further.  \pyro\ provides an easy
test bed for experimenting with these ideas.

To make the code simple, we restrict ourselves to pure V-cycles and
solve on square domains with the number of zones a power of 2.
Red-black Gauss-Seidel relaxation is done.  The restriction operation
is simply averaging the four fine cells into the corresponding coarse
cell.  For prolongation, we create centered-slopes in each direction
centered for every coarse cell and use a bilinear interpolation
(without the cross $xy$ term) to fill each of the four finer cells.
Furthermore, we only support homogeneous Dirichlet or Neumann boundary
conditions.  The grid is coarsened until we get down to a $2^2$ grid,
at which point the residual equation is solved by pure relaxation,
doing, by default, 50 smoothing iterations on the $2^2$ grid.

For homogeneous Dirichlet boundary conditions, the goal is to have the
solution be 0 exactly at the interface of the zone touching the domain
boundary.  Consider the left boundary (see Figure~\ref{fig:grids}): we want
$\phi_a = 0$.  The standard approach (to second-order accuracy)
is to average the adjacent cells to that interface, and set this to zero,
giving:
\begin{equation}
\phi_{\mathrm{lo}-1,j} = -\phi_{\mathrm{lo},j}
\end{equation}
A similar construction is done at the upper boundary in $x$, and for the
boundaries in $y$.  Neumann BCs are done in a similar fashion---we construct
a difference that is centered at the interface, so to second-order, we have
\begin{equation}
\phi_{\mathrm{lo}-1,j} = \phi_{\mathrm{lo},j}
\end{equation}
Inhomogeneous boundary conditions would be treated similarly, but now
the boundary value itself will appear in the expression for the ghost
cell.

\subsubsection{Poisson-solve test problem}

To test the multigrid solver, we solve a Poisson problem with a known analytic
solution.  This example comes from \citet{mgtutorial}:
\begin{align}
\nabla^2 \phi = - 2 \bigl [& (1-6x^2)y^2(1-y^2) +  \nonumber \\
                           & (1-6y^2)x^2(1-x^2) \bigr ]
\end{align}
solved on a unit square with $\phi = 0$ on the boundary.  The analytic solution
in this case is:
\begin{equation}
\phi(x,y) = (x^2 - x^4)(y^4 - y^2)
\end{equation}
We continue to cycle until the relative error (L2 norm of the residual
/ L2 norm of the source) is less than $10^{-11}$.  Figure~\ref{fig:mg}
shows L2-norm of the absolute error of our solution compared to the
analytic solution for a variety of resolutions.  We see perfect
second-order convergence.

\subsubsection{Explorations for students}

\begin{itemize}
    \addtolength{\itemsep}{-0.5\baselineskip}
\item Instead of doing multigrid, run with smoothing only and look
  at how long it takes to converge.

\item Implement inhomogeneous BCs.

\item Experiment with different bottom solvers.
\end{itemize}

\subsection{Implicit diffusion}

Many phenomena can be described by the diffusion equation:
\begin{equation}
\phi_t = k \phi_{xx}
\end{equation}
including thermal diffusion/conduction and viscosity.  Here, $k$ is
the diffusion coefficient, which we will take to be constant, and
$\phi$ is the scalar quantity being diffused.  The diffusion solver
uses multigrid to solve an implicit Crank-Nicolson (centered in time) discretization of the diffusion
equation:
\begin{equation}
\frac{\phi^{n+1}_i - \phi^n_i}{\Delta t} =
   \frac{1}{2} \left ( k \nabla^2 \phi^n_i + k \nabla^2 \phi^{n+1}_i \right )
\end{equation}
Grouping all the $n+1$ terms on the left, we find:
\begin{equation}
\phi^{n+1}_i - \frac{\Delta t}{2} k \nabla^2 \phi^{n+1}_i =
    \phi^n_i + \frac{\Delta t}{2} k \nabla^2 \phi^n_i
\end{equation}
This is in the form of a constant-coefficient Helmholtz equation,
Eq.~\ref{eq:helmholtz}, with
\begin{equation}
\alpha = 1, \,\, \beta = \frac{\Delta t}{2} k , \,\,
f = \phi^n + \frac{\Delta t}{2} k \nabla^2 \phi^n
\end{equation}
An update over a single timestep is achieved by simply calling the
multigrid solver.  Since this solver is implicit, there is no timestep
limit for stability, but accuracy will of course be better with a
smaller timestep.  For this solver, the CFL number in the driver, $C$,
is based on the explicit timestep limit:
\begin{equation}
\Delta t = C \frac{\Delta x^2}{k}
\end{equation}
where $C \le 1/2$ is needed for a standard explicit discretization of
the diffusion equation.

\subsubsection{Diffusion of a Gaussian test problem}

\begin{figure}
\centering
\includegraphics[width=0.9\linewidth]{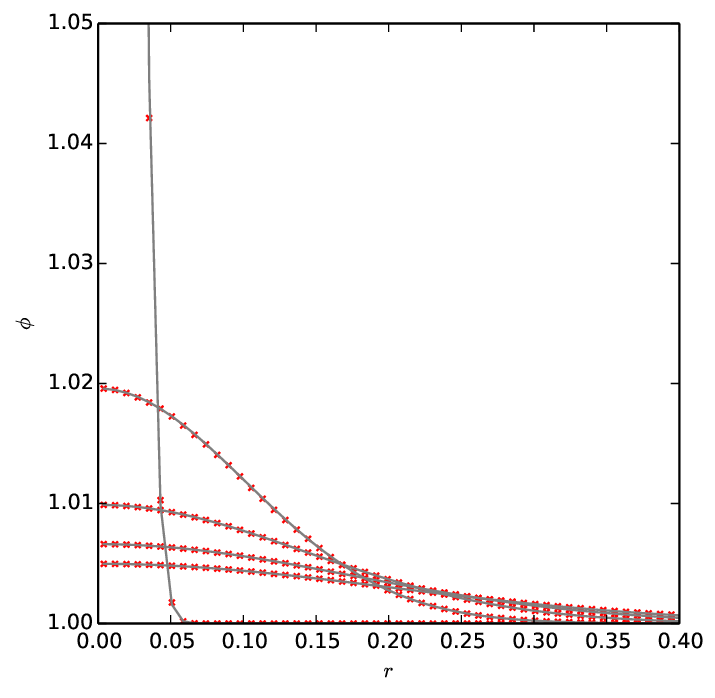}
\caption{\label{fig:diff} Radial profile of the diffusion of a
  Gaussian shown every $\Delta t = 0.005$ starting with $t = 0$,
  using $C = 2.0$ and a $128^2$ grid.
  The analytic solution is shown as the gray line.  We see excellent
  agreement.  This example was run with: {\tt ./pyro.py diffusion gaussian inputs.gaussian}, and the comparison with the analytic solution was done
  with the {\tt analysis/gauss\_diffusion\_compare.py} script.}
\end{figure}
An initial Gaussian profile remains a Gaussian under the action of
diffusion, with the amplitude decreasing and width increasing.  This
allows us to test the diffusion solver against the analytic solution
(see, e.g.~\citealt{swestymyra:2009}):
\begin{equation}
   \phi(x,t) = (\phi_2 - \phi_1) {\frac{t_0}{t + t_0}} e^{-\frac{1}{4}[(x - x_c)^2 + (y-y_c)^2]/k(t+t_0)} + \phi_1
  \end{equation}
For the initial conditions, we take $t_0 = 0.0001$, $\phi_1 = 1$, and
$\phi_2 = 2$.  We run with $k = 1$.  Figure~\ref{fig:diff} shows the
angle-averaged radial profile at several different times together with
the analytic solution for a run on a $128^2$ grid with $C = 2.0$.  We see
excellent agreement.

\subsubsection{Exercises for students}

There are a number of straightforward exercises and extensions
\begin{itemize}
    \addtolength{\itemsep}{-0.5\baselineskip}
\item Experiment with different-sized timesteps and initially
  discontinuous data.
\item Implement a non-constant diffusion coefficient solver.
\item Compare Crank-Nicolson to backwards Euler.
\end{itemize}

\subsection{Incompressible Hydrodynamics}

The equations of incompressible flow are:
\begin{align}
U_t + U \cdot \nabla U + \nabla p &= 0 \\
\nabla \cdot U &= 0
\end{align}
Incompressible flow adds a additional complexity to our systems---now
an elliptic constraint is present on the velocity field that must also
be satisfied at each timestep.  It is also an important stepping-stone
toward understanding low Mach number methods, used for both smallscale
combustion \citep{Bell:2004} and stratified flows \citep{multilevel}.

\pyro's incompressible solver follows a second-order projection
methodology (see, \citealt{chorin:1968,BCG}).
A projection method relies on the fact that any vector field, $U^\star$
can be decomposed into a divergence-free part, $U^d$, and the gradient
of a scalar:
\begin{equation}
U^\star = U^d + \nabla \phi
\label{eq:decomp}
\end{equation}
The idea is that we first use the same unsplit advection techniques as
with linear advection and compressible flow to update $U$ to the new
time, giving a velocity field $U^\star$ that does not yet satisfy the
divergence constraint.  By taking the divergence of
Eq.~\ref{eq:decomp}, we get a Poisson equation for the scalar $\phi$
needed to correct our velocity field and make it divergence free:
\begin{equation}
\nabla^2 \phi = \nabla \cdot U^\star
\end{equation}
This is then solved using multigrid, resulting in the new divergence-free
velocity field.

There are a lot of variations on this idea.  First, approximate
projections make use of discretizations of the divergence, $D$, and
gradient, $G$, operators that together are not the same as the
discretized Laplacian, $L$ (i.e. $DG\phi \ne L\phi$).  This
approximation however can result in a more robust discretization.
Finally, some methods put $\phi$ at the nodes of the cells while
others make it cell-centered.  We choose the latter here, as a
cell-centered discretization allows us to reuse our existing multigrid
solver.  An additional complexity is that a projection is also done
on the predicted half-time, interface velocities that are used
to construct the flux---this is needed for stability to 
CFL numbers of unity~\citep{BCH}.

The implementation in \pyro\ is pieced together from a variety of
sources.  \citet{BCH} describes a cell-centered method, but with an
exact projection (with a larger, decoupled stencil).  \citet{ABS}
describes an approximate projection method, but with a node-centered
final projection.  We follow this paper closely up until the
projection.  We then do the cell-centered projection described in
\citet{MartinColella} (and Martin's PhD thesis).  All of these method
are largely alike, aside from how the discretization of the final
projection is handled.  The advective part follows the CTU methodology
from the advection solver very closely (but the Riemann solver is now
that of Burger's equation instead of the linear advection equation).

\subsubsection{Convergence test}

\begin{figure}
\centering
\includegraphics[width=0.9\linewidth]{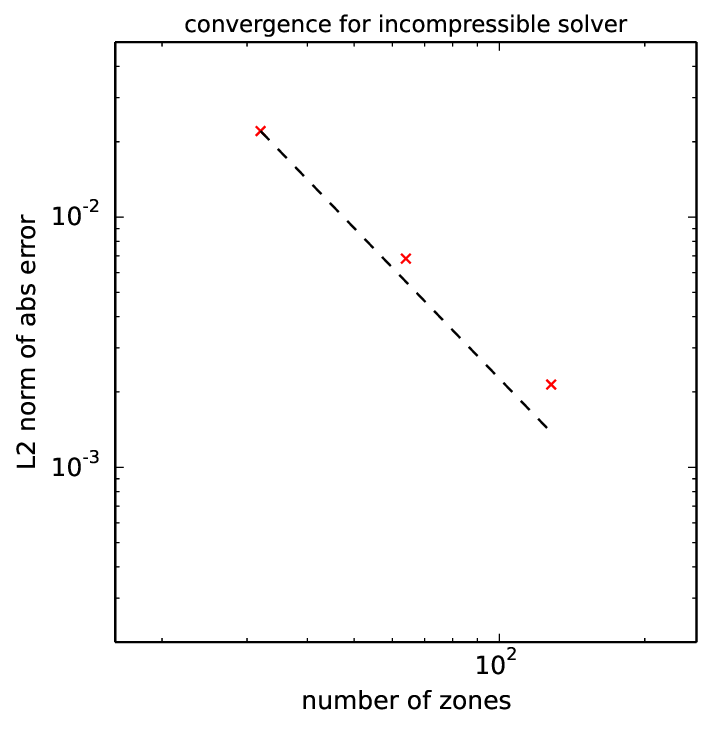}
\caption{\label{fig:incomp_converge} Convergence of the incompressible
  solver.  These tests were run with {\tt ./pyro.py incompressible
    converge inputs.converge.32}, and then with the {\tt
    inputs.converge.64} and {\tt inputs.converge.128} inputs files.
    The error with respect to the analytic solution was computed using
    the {\tt analysis/incomp\_converge\_error.py} script.}
\end{figure}

A standard convergence test for incompressible flow was described by
\citet{minion:1996}.  There, the initial velocity field is:
\begin{align}
u(x,y) &= 1 - 2 \cos(2 \pi x) \sin(2 \pi y)  \\
v(x,y) &= 1 + 2 \sin(2 \pi x) \cos(2 \pi y)
\end{align}
and the solution at a later time is 
\begin{align}
u(x,y,t) &= 1 - 2 \cos(2 \pi (x - t)) \sin(2 \pi (y - t)) \\
v(x,y,t) &= 1 + 2 \sin(2 \pi (x - t)) \cos(2 \pi (y - t)) 
\end{align}
By comparing the numerical solution to the analytic solution, we can 
compute the error.  We use a fixed $\Delta t/\Delta x$ and run at 
several resolutions.  Figure~\ref{fig:incomp_converge} shows the results.
We see nearly second-order convergence---the departure is due to the use
of the limiters.

\subsubsection{Explorations for students}

\begin{itemize}
    \addtolength{\itemsep}{-0.5\baselineskip}
\item Add viscosity to the system. This will require doing 2 parabolic
  solves (one for each velocity component). These solves will look
  like the diffusion operation, and will update the provisional
  velocity field.

\item Switch to a variable density system~\citep{BellMarcus}. This
  will require adding a mass continuity equation that is advected and
  switching the projections to a variable-coefficient form (since
  $\rho$ now enters).
\end{itemize}

\section{Summary}

\pyro\ provides the building blocks that together make up a modern
astrophysical simulation code.  By providing simple solvers in a
python environment, students are encouraged to experiment.  We showed
that the core solvers perform as expected and motivated some simple
(and some not-so-simple) extensions for each of the solvers.

The core algorithms in \pyro\ are mature and provide a good basis for
new students to learn the ins-and-outs of computational hydrodynamics
through self-study.
The code will continue to be maintained and keep up with new
developments in python and software engineering, but we do not expect
to add new intricacies to the existing solvers.  The software engineering
aspects built into \pyro\ can provide a basis for learning
about version control, regression testing, and verification and
validation---at the very least by showing
them these ideas exist and a giving them the ability to explore.

The code is freely available, and this paper together with the code,
and the detailed derivations on the \pyro\ website provide the
guidance for students.  The code and online
notes are written for students to learn on their own, or as part
of a class.  Furthermore, a mailing list is provided to
support users of the code\footnote{\url{http://bender.astro.sunysb.edu/mailman/listinfo/pyro-help}}.

Major changes in the future will focus on new systems of PDEs.  We
envision solvers for MHD, radiation hydrodynamcs in the flux-limited
diffusion approximation, and some examples of multiphysics (like
diffusion + reaction systems and the viscous Burger's equation).  These
can be integrated into the same framework present now and will help
introduce new ideas to the students.  We will also focus on making \pyro\
more of a laboratory to introduce students to software development practices
and add unit testing to the various modules.  Finally, we hope to 
develop some simple 1-d examples for each solver that can be run in
IPython\footnote{\url{http://ipython.org}} notebooks to help futher link \pyro\ and the online notes.

\section*{Acknowledgements} The results shown here can be
reproduced with the \pyro\ git version: 3a1794743255d6f748a61dc0841576b9c8c54d3b.
We thank Ann Almgren, John Bell, Alan Calder, 
Chris Malone, and Andy Nonaka for many helpful discussions on
hydrodynamics and/or this paper.  We thank the anonymous referee
for very helpful suggestions and a thorough review of the code.  Much
of the effort in making the code conform to the python style guidelines
is due to this feedback.  \pyro\ was developed over many years
as a tool to learn these methods for my own research before
incorporating them into production codes.  The original version of
\pyro\ benefited greatly from the inviting environment at Lulu
Carpenter's.  Some support along the way was provided by DOE Office of
Nuclear Physics grant DOE DE-FG02-06ER41448 and NSF grant
AST-1211563. \\

\bibliographystyle{elsarticle-harv}

\begin{thebibliography}{35}
\expandafter\ifx\csname natexlab\endcsname\relax\def\natexlab#1{#1}\fi
\expandafter\ifx\csname url\endcsname\relax
  \def\url#1{\texttt{#1}}\fi
\expandafter\ifx\csname urlprefix\endcsname\relax\def\urlprefix{URL }\fi

\bibitem[{{Agertz} et~al.(2007){Agertz}, {Moore}, {Stadel}, {Potter},
  {Miniati}, {Read}, {Mayer}, {Gawryszczak}, {Kravtsov}, {Nordlund}, {Pearce},
  {Quilis}, {Rudd}, {Springel}, {Stone}, {Tasker}, {Teyssier}, {Wadsley}, and
  {Walder}}]{agertz:2007}
{Agertz}, O., {Moore}, B., {Stadel}, J., {Potter}, D., {Miniati}, F., {Read},
  J., {Mayer}, L., {Gawryszczak}, A., {Kravtsov}, A., {Nordlund}, {\AA}.,
  {Pearce}, F., {Quilis}, V., {Rudd}, D., {Springel}, V., {Stone}, J.,
  {Tasker}, E., {Teyssier}, R., {Wadsley}, J., {Walder}, R., Sep. 2007.
  {Fundamental differences between SPH and grid methods}. Monthly Notices of
  the Royal Astronomical Society 380, 963--978.

\bibitem[{{Almgren} et~al.(2010){Almgren}, {Beckner}, {Bell}, {Day}, {Howell},
  {Joggerst}, {Lijewski}, {Nonaka}, {Singer}, and {Zingale}}]{almgren:2010}
{Almgren}, A.~S., {Beckner}, V.~E., {Bell}, J.~B., {Day}, M.~S., {Howell},
  L.~H., {Joggerst}, C.~C., {Lijewski}, M.~J., {Nonaka}, A., {Singer}, M.,
  {Zingale}, M., Jun. 2010. {CASTRO: A New Compressible Astrophysical Solver.
  I. Hydrodynamics and Self-gravity}. Astrophys J 715, 1221--1238.

\bibitem[{Almgren et~al.(1996)Almgren, Bell, and Szymczak}]{ABS}
Almgren, A.~S., Bell, J.~B., Szymczak, W.~G., Mar. 1996. A numerical method for
  the incompressible {N}avier-{S}tokes equations based on an approximate
  projection. SIAM J. Sci. Comput. 17~(2), 358--369.

\bibitem[{{Bell} et~al.(1989){Bell}, {Colella}, and {Glaz}}]{BCG}
{Bell}, J.~B., {Colella}, P., {Glaz}, H.~M., Dec. 1989. {A Second Order
  Projection Method for the Incompressible Navier-Stokes Equations}. Journal of
  Computational Physics 85, 257.

\bibitem[{Bell et~al.(1991)Bell, Colella, and Howell}]{BCH}
Bell, J.~B., Colella, P., Howell, L.~H., Jun. 1991. An efficient second-order
  projection method for viscous incompressible flow. In: Proceedings of the
  Tenth AIAA Computational Fluid Dynamics Conference. AIAA, pp. 360--367, see
  also: https://seesar.lbl.gov/anag/publications/colella/A\_2\_10.pdf.

\bibitem[{Bell et~al.(1988)Bell, Dawson, and Shubin}]{BDS}
Bell, J.~B., Dawson, C.~N., Shubin, G.~R., 1988. An unsplit, higher order
  {G}odunov method for scalar conservation laws in multiple dimensions. Journal
  of Computational Physics 74, 1--24.

\bibitem[{Bell et~al.(2004)Bell, Day, Rendleman, Woosley, and
  Zingale}]{Bell:2004}
Bell, J.~B., Day, M.~S., Rendleman, C.~A., Woosley, S.~E., Zingale, M.~A.,
  2004. Adaptive low {M}ach number simulations of nuclear flame microphysics.
  Journal of Computational Physics 195~(2), 677--694.

\bibitem[{Bell and Marcus(1992)}]{BellMarcus}
Bell, J.~B., Marcus, D.~L., 1992. A second-order projection method for
  variable-density flows. Journal of Computational Physics 101~(2), 334 -- 348.

\bibitem[{Briggs et~al.(2000)Briggs, Henson, and McCormick}]{mgtutorial}
Briggs, W.~L., Henson, V.-E., McCormick, S.~F., 2000. A Multigrid Tutorial, 2nd
  Ed. SIAM.

\bibitem[{Chorin(1968)}]{chorin:1968}
Chorin, A.~J., 1968. Numerical solution of the {N}avier-{S}tokes equations.
  Math. Comp. 22, 745--762.

\bibitem[{{Colella}(1985)}]{colella:1985}
{Colella}, P., 1985. {A direct Eulerian MUSCL scheme for gas dynamics}. SIAM J
  Sci Stat Comput 6~(1), 104--117.

\bibitem[{{Colella}(1990)}]{colella:1990}
{Colella}, P., Mar. 1990. {Multidimensional upwind methods for hyperbolic
  conservation laws}. Journal of Computational Physics 87, 171--200.

\bibitem[{{Colella} and {Woodward}(1984)}]{colellawoodward:1984}
{Colella}, P., {Woodward}, P.~R., Sep. 1984. {The Piecewise Parabolic Method
  (PPM) for Gas-Dynamical Simulations}. Journal of Computational Physics 54,
  174--201.

\bibitem[{{Ferland}(2002)}]{ferland:2002}
{Ferland}, G., Oct. 2002. {Reliability in the Face of Complexity; The Challenge
  of High-End Scientific Computing}. ArXiv Astrophysics e-prints.

\bibitem[{{Flash Center for Computational Science}(2014)}]{flashdoc}
{Flash Center for Computational Science}, 2014. Flash user's guide, version
  4.2~\url{http://flash.uchicago.edu/site/flashcode/user_support/flash4_ug_4p2/}.

\bibitem[{{Fryxell} et~al.(2000){Fryxell}, {Olson}, {Ricker}, {Timmes},
  {Zingale}, {Lamb}, {MacNeice}, {Rosner}, {Truran}, and {Tufo}}]{flash}
{Fryxell}, B., {Olson}, K., {Ricker}, P., {Timmes}, F.~X., {Zingale}, M.,
  {Lamb}, D.~Q., {MacNeice}, P., {Rosner}, R., {Truran}, J.~W., {Tufo}, H.,
  Nov. 2000. {FLASH}: An adaptive mesh hydrodynamics code for modeling
  astrophysical thermonuclear flashes. Astrophysical Journal Supplement 131,
  273--334.

\bibitem[{{Hubber} et~al.(2013){Hubber}, {Falle}, and {Goodwin}}]{hubber:2013}
{Hubber}, D.~A., {Falle}, S.~A.~E.~G., {Goodwin}, S.~P., Jun. 2013.
  {Convergence of AMR and SPH simulations - I. Hydrodynamical resolution and
  convergence tests}. Monthly Notices of the Royal Astronomical Society 432,
  711--727.

\bibitem[{{Kamm} and {Timmes}(2007)}]{timmes_sedov_code}
{Kamm}, J.~R., {Timmes}, F.~X., 2007Submitted to ApJ supplement, May 2007, see
  \texttt{http://cococubed.asu.edu/code\_pages/sedov.shtml}.

\bibitem[{LeVeque(2002)}]{leveque:2002}
LeVeque, R.~J., 2002. Finite-Volume Methods for Hyperbolic Problems. Cambridge
  University Press.

\bibitem[{{Martin} and {Colella}(2000)}]{MartinColella}
{Martin}, D.~F., {Colella}, P., Sep. 2000. {A Cell-Centered Adaptive Projection
  Method for the Incompressible Euler Equations}. Journal of Computational
  Physics 163, 271--312.

\bibitem[{{Miller} and {Colella}(2002)}]{millercolella:2002}
{Miller}, G.~H., {Colella}, P., Nov. 2002. {A Conservative Three-Dimensional
  Eulerian Method for Coupled Solid-Fluid Shock Capturing}. Journal of
  Computational Physics 183, 26--82.

\bibitem[{{Minion}(1996)}]{minion:1996}
{Minion}, M.~L., 1996. {A Projection Method for Locally Refined Grids}. Journal
  of Computational Physics 127, 158--177.

\bibitem[{Nonaka et~al.(2010)Nonaka, Almgren, Bell, Lijewski, Malone, and
  Zingale}]{multilevel}
Nonaka, A., Almgren, A.~S., Bell, J.~B., Lijewski, M.~J., Malone, C.~M.,
  Zingale, M.~., 2010. {\tt MAESTRO}:an adaptive low mach number hydrodynamics
  algorithm for stellar flows. Astrophysical Journal Supplement 188, 358--383.

\bibitem[{{Schulz-Rinne} et~al.(1993){Schulz-Rinne}, {Collins}, and
  {Glaz}}]{schulz-rinne:1993}
{Schulz-Rinne}, C.~W., {Collins}, J.~P., {Glaz}, H.~M., 1993. {Numerical
  Solution of the Riemann Problem for Two-Dimensional Gas Dynamics}. SIAM J Sci
  Comput 14~(6), 1394--1414.

\bibitem[{{Sedov}(1959)}]{sedov:1959}
{Sedov}, L.~I., 1959. Similarity and Dimensional Methods in Mechanics. Academic
  Press, translated from the 4th Russian Ed.

\bibitem[{{Shamir} et~al.(2013){Shamir}, {Wallin}, {Allen}, {Berriman},
  {Teuben}, {Nemiroff}, {Mink}, {Hanisch}, and {DuPrie}}]{shamir:2013}
{Shamir}, L., {Wallin}, J.~F., {Allen}, A., {Berriman}, B., {Teuben}, P.,
  {Nemiroff}, R.~J., {Mink}, J., {Hanisch}, R.~J., {DuPrie}, K., Feb. 2013.
  {Practices in source code sharing in astrophysics}. Astronomy and Computing
  1, 54--58.

\bibitem[{{Sod}(1978)}]{sod:1978}
{Sod}, G.~A., Apr. 1978. {A survey of several finite difference methods for
  systems of nonlinear hyperbolic conservation laws}. Journal of Computational
  Physics 27, 1--31.

\bibitem[{{Stone} et~al.(2008){Stone}, {Gardiner}, {Teuben}, {Hawley}, and
  {Simon}}]{athena}
{Stone}, J.~M., {Gardiner}, T.~A., {Teuben}, P., {Hawley}, J.~F., {Simon},
  J.~B., Sep. 2008. {Athena: A New Code for Astrophysical MHD}. Astrophys J
  Suppl S 178, 137--177.

\bibitem[{{Strang}(1968)}]{strang:1968}
{Strang}, G., 1968. {On the construction and comparison of difference schemes}.
  {SIAM J. Numerical Analysis} 5, 506--517.

\bibitem[{{Swesty} and {Myra}(2009)}]{swestymyra:2009}
{Swesty}, F.~D., {Myra}, E.~S., Mar. 2009. {A Numerical Algorithm for Modeling
  Multigroup Neutrino-Radiation Hydrodynamics in Two Spatial Dimensions}.
  Astrophysical Journal Supplement 181, 1--52.

\bibitem[{Toro(1997)}]{toro:1997}
Toro, E.~F., 1997. Riemann Solvers and Numerical Methods for Fluid Dynamics.
  Springer.

\bibitem[{{Turk}(2013)}]{turk:2013}
{Turk}, M.~J., Jan. 2013. {How to Scale a Code in the Human Dimension}. ArXiv
  e-prints.

\bibitem[{{Turk} et~al.(2011){Turk}, {Smith}, {Oishi}, {Skory}, {Skillman},
  {Abel}, and {Norman}}]{yt}
{Turk}, M.~J., {Smith}, B.~D., {Oishi}, J.~S., {Skory}, S., {Skillman}, S.~W.,
  {Abel}, T., {Norman}, M.~L., Jan. 2011. {yt: A Multi-code Analysis Toolkit
  for Astrophysical Simulation Data}. Astrophysical Journal Supplement 192, 9.

\bibitem[{{Wilson} et~al.(2012){Wilson}, {Aruliah}, {Titus Brown}, {Chue Hong},
  {Davis}, {Guy}, {Haddock}, {Huff}, {Mitchell}, {Plumbley}, {Waugh}, {White},
  and {Wilson}}]{wilson:2012}
{Wilson}, G., {Aruliah}, D.~A., {Titus Brown}, C., {Chue Hong}, N.~P., {Davis},
  M., {Guy}, R.~T., {Haddock}, S.~H.~D., {Huff}, K., {Mitchell}, I.~M.,
  {Plumbley}, M., {Waugh}, B., {White}, E.~P., {Wilson}, P., Sep. 2012. {Best
  Practices for Scientific Computing}. ArXiv e-prints.

\bibitem[{{Zingale} et~al.(2002){Zingale}, {Dursi}, {ZuHone}, {Calder},
  {Fryxell}, {Plewa}, {Truran}, {Caceres}, {Olson}, {Ricker}, {Riley},
  {Rosner}, {Siegel}, {Timmes}, and {Vladimirova}}]{hse}
{Zingale}, M., {Dursi}, L.~J., {ZuHone}, J., {Calder}, A.~C., {Fryxell}, B.,
  {Plewa}, T., {Truran}, J.~W., {Caceres}, A., {Olson}, K., {Ricker}, P.~M.,
  {Riley}, K., {Rosner}, R., {Siegel}, A., {Timmes}, F.~X., {Vladimirova}, N.,
  Dec. 2002. {Mapping Initial Hydrostatic Models in Godunov Codes}. Astrophys J
  Suppl S 143, 539--565.

\end{thebibliography}

\end{document}